\documentclass[prb,aps,preprint,superscriptaddress]{revtex4-1}
\usepackage[dvips]{graphics}
\begin{document}
\title{Magnetic ground state of distorted 6$H$ perovskite Ba$_3$CdIr$_2$O$_9$}
\author{Md Salman Khan}
\affiliation{School of Materials Science, Indian Association for the Cultivation of Science, 2A \& 2B Raja S. C. Mullick Road, Jadavpur, Kolkata 700032, India}
\author{Abhisek Bandyopadhyay}
\affiliation{School of Materials Science, Indian Association for the Cultivation of Science, 2A \& 2B Raja S. C. Mullick Road, Jadavpur, Kolkata 700032, India}
\author{Abhishek Nag}
\affiliation{School of Materials Science, Indian Association for the Cultivation of Science, 2A \& 2B Raja S. C. Mullick Road, Jadavpur, Kolkata 700032, India}
\author{Vinod Kumar}
\affiliation{Department of  Physics, Indian Institute Of Technology Bombay, Powai, Mumbai 400076, India}
\author{A. V. Mahajan}
\affiliation{Department of  Physics, Indian Institute Of Technology Bombay, Powai, Mumbai 400076, India}
\author{Sugata Ray}
\email[Corresponding author:] {mssr@iacs.res.in}
\affiliation{School of Materials Science, Indian Association for the Cultivation of Science, 2A \& 2B Raja S. C. Mullick Road, Jadavpur, Kolkata 700032, India}
\affiliation{Technical Research Center, Indian Association for the Cultivation of Science, 2A \& 2B Raja S. C. Mullick Road, Jadavpur, Kolkata 700032, India}

\date{\today}

\begin{abstract}
Perovskite iridates of 6$H$ hexagonal structure present a plethora of possibilities in terms of the variety of ground states resulting from a competition between spin-orbit coupling (SOC), hopping, noncubic crystal field ($\Delta^{\text{NC}}_{\text{CFE}}$) and superexchange energy scales within the Ir$_2$O$_9$ dimers. Here we have investigated one such compound Ba$_3$CdIr$_2$O$_9$ by x-ray diffraction, $dc$ magnetic susceptibility($\chi$), heat capacity($C_p$) and also ${}^{113}$Cd nuclear magnetic resonance (NMR) spectroscopy. We have established that the magnetic ground state has a small but finite magnetic moment on Ir$^{5+}$ in this system, which likely arises from intradimer Ir-Ir hopping and local crystal distortions. Our heat capacity, NMR, and $dc$ magnetic susceptibility measurements further rule out any kind of magnetic long-/short range ordering among the Ir moments down to at least 2K . In addition, the magnetic heat capacity data shows linear temperature dependence at low temperatures under applied high fields ($>$ 30 kOe), suggesting gapless spin-density of states in the compound.
\end{abstract}

\maketitle

\newpage

\section{introduction}
5$d$ transition metal oxides, especially iridates, have become a fertile ground for hosting diverse spectrum of exotic physical properties due to the intricate competition between spin-orbit coupling (SOC), crystal field energy ($\Delta_{\text{CFE}}$), and on-site Coulomb repulsion ($U$) ~\cite{Krempa,Lawler,Pesin, Matsuhira,Fukazawa,Nakatsuji,Kim1,Moon,Kim2}. In contrast to the traditional wisdom of uncorrelated band metallicity in iridates due to extended 5$d$ orbitals, B.J.Kim {\it et al.} first established the central role of SOC on the Mott insulating ground state in the layered tetravalent iridate Sr$_2$IrO$_4$~\cite{Kim1}. Extension of the same idea to pentavalent iridates (Ir$^{5+}$: 5$d^4$) gives rise to different situations. Substantial SOC ($\lambda$) in Ir$^{5+}$ produces ${}^6C_4$ = 15 possible $J$ states or organisations of electrons within the 6 $t_{2g}$ orbitals, with $J$ = 0 as the ground state~\cite{Khluin}. But a purely nonmagnetic $J$ = 0 state has not been encountered till date in any of the reported 5$d^4$ iridate systems~\cite{Cao,Marco,Dey,Holm}. It has been speculated that the magnetism in these $d^4$ compounds may arise from the Van Vleck-type intrasite singlet-triplet excitations ($J$ = 0 $\rightarrow$ $J$ = 1) when the superexchange (mediated by complex Ir-O-Ir paths) energy scale of 4$t^2$/2$U$ becomes comparable to SOC-induced singlet-triplet energy gap, allowing virtual transitions to higher levels and magnetic condensation of Van Vleck excitons~\cite{Khluin}. On the other hand, solid state effects such as large bandwidth of the 5$d$ orbitals, ligand-Ir charge transfer, non-cubic crystal field and intersite Ir-Ir hopping always act against the atomic SOC effect. Moreover, Ir-Ir hopping can modify the ground state itself by creating several mixed $J$ states and a finite moment for which excitonic mechanisms are not relevant~\cite{Nagprb2}. Of course the other possibility is that these all effectively reduce the SOC strength ($\lambda$) on the Ir-site and as a result, spontaneous magnetic moments may be generated~\cite{Meetei,Chen,Liu,S.B,Kim3,Dodds,Jackeli,Khomski}. The closest realization of non-magnetic $J = 0$ ground state in the $d^4$ systems till date is recorded in the 6$H$ hexagonal perovskite compound Ba$_3$ZnIr$_2$O$_9$ where a tiny moment of 0.2$\mu_B$/Ir appears on each Ir$^{5+}$ site due to strong intradimer hopping mechanism within a Ir$_2$O$_9$ dimer unit~\cite{Nagprl}.
\par
Here, in this paper we report a new compound Ba$_3$CdIr$_2$O$_9$ (BCIO) which belongs to the family of 6$H$ hexagonal perovskite Ba$_3$$M$Ir$_2$O$_9$ ($M$ = akaline earth or rare earth element). Although rigorous experimental studies have been performed on the 6$H$ Ba$_3$$M$Ir$_2$O$_9$ compounds with $M$= Zn, Mg, Ca, Sr, Y, Sc and In~\cite{Nagprb,Nagprl,Dey3,Dey4,Sakomoto}, showing diverse ground state properties, the $M$ = Cd member has not been carefully looked at, except an earlier report describing only the synthesis and structural details of the compound~\cite{Sakomoto}.
\par
The standard 6$H$ hexagonal structure consists of a single $M$O$_6$ octahedron and  Ir$_2$O$_9$ dimers of two face sharing IrO$_6$ octahedra [Fig.1(e)]. The Ir-Ir structural dimers are constructed parallel to the crystallographic $c$-axis and these dimers are linked with each other to form an edge-shared triangular lattice of Ir atoms in the $a-b$ plane [Fig.1(c)]. Such geometries next dictate how the tiny Ir spins should interact with each other which is an outcome of different Ir-Ir exchange interactions ~\cite{Nagprb}, as displayed by blue curved arrows in Fig. 1(b) and 1(c) and by blue dashed lines in Fig. 1(d). Also, the local octahedral distortions influencing the Ir-O bond  lengths and Ir-O-Ir bond angles further complicate  the magnetic ground state in this system. A. Nag {\it et al.}~\cite{Nagprb} has shown that the competition between SOC and octahedral distortions in Ba$_3M$Ir$_2$O$_9$ ($M$ = Mg, Ca, Sr) series of compounds greatly influences the magnetic properties. Due to similar ionic radii of Mg and Zn, the $M$=Mg compound bears closer analogy to the spin orbital liquid Ba$_3$ZnIr$_2$O$_9$ ~\cite{Nagprl}, while $M$ = Ca and Sr compounds, due to higher degree of monoclinic ($C2/c$) structural distortions, reveal differences in magnetic behaviour. Although the amount of octahedral distortion was not monotonic with the ionic radius of $M$ cations, interdimer Ir-Ir distance increases and intradimer Ir-Ir distance decreases regularly with increasing $M^{2+}$ cation size~\cite{Nagprb}. The 6$H$-hexagonal perovskite Ba$_3$CdIr$_2$O$_9$ becomes particularly important in this context where Cd$^{2+}$ has ionic radius close to Ca ($r_{Ca^{2+}}$ = 1.00 \AA) but in between the Mg ($r_{Mg^{2+}}$ = 0.72 \AA) and Sr($r_{Sr^{2+}}$ = 1.18 \AA).
\par
Another salient notion for looking at the Ba$_3$CdIr$_2$O$_9$ compound is that unlike $M$ = Zn, Mg, Ca and Sr compounds, the Cd counterpart has an NMR active ${}^{113}$Cd ($I$ = 1/2) nucleus with reasonable natural abundance. Therefore, ${}^{113}$Cd NMR becomes an exciting option to explore the local spin dynamics around Ir in this compound. Keeping this in mind, here in this paper we have investigated the 6$H$ Ba$_3$CdIr$_2$O$_9$ compound using standard X-ray diffraction (XRD), X-ray photoelectron spectroscopy (XPS), electric resistivity, $dc$ magnetic susceptibility, heat capacity,  and ${}^{113}$Cd NMR measurements.  Our combined $dc$ susceptibility and NMR studies confirm the true intrinsic nature of the developed tiny Ir-moments ($\mu_{eff}$ = 0.3 $\mu_B$/Ir) in this compound. Despite having considerable antiferromagntic (AFM) interaction ($\Theta_{CW}$ = -21 K) between these Ir$^{5+}$ moments, the Ba$_3$CdIr$_2$O$_9$ compound does not order down to at least 2 K (frustration parameter $>$ 10), due to geometric frustration arising from the triangular lattice [Fig. 1(c)-(d)], which is markedly different from $M$ = Ca, Sr compounds. Further, the magnetic heat capacity reveals a linear temperature dependence at low-$T$ unfolding the gapless nature of spin excitations within this compound.

\section{Experimental Methods}
Polycrystalline samples of Ba$_3$CdIr$_2$O$_9$ were prepared by standard solid state reaction methods using high purity ($>$99.9\%) starting materials BaCO$_3$, CdO, IrO$_2$ (from Sigma Aldrich) in an appropriate ratio. These mixtures were thoroughly ground and pressed into pellets before initial calcination at 800$^{\circ}$C for 12h. Finally the as-calcined pellet was annealed at 1100$^{\circ}$C four times for 12 hours each in air with several intermediate grindings. The phase purity of the sample was checked from X-ray powder diffraction measured at Rigaku Smartlab x-ray diffractometer with Cu $K_{\alpha}$ radiation at room temperature as well as low temperatures. The crystal structure of this sample was obtained after refining the XRD data by Rietveld technique using FULLPROF program~\cite{Carvajal}. To check homogeneity and stoichiometry in the sample, Energy Dispersive X-Ray Analysis (EDX) was also performed using Field Emission Scanning Electron Microscope (FE-SEM, JEOL, JSM-7500F). Temperature dependent electrical resistivity was measured using the four probe technique  in a laboratory based experimental set up. The X-ray photoelectron spectroscopy (XPS) measurements were carried out using OMICRON electron spectrometer, equipped with SCIENTA OMICRON SPHERA analyzer and Al $K_{\alpha}$ monochromatic source with an energy resolution of 0.5 eV. The sample surface has been cleaned before experiment by {\it in-situ} Ar-sputtering. The collected spectra were then processed and analyzed with Kolxpd program. The temperature and magnetic field dependent $dc$ magnetization was carried out using the Quantum Design (SQUID) magnetometer. Heat capacity both in zero field and several higher magnetic fields (upto 9 Tesla) were measured using the heat capacity attachment of a Quantum Design PPMS. Measurements of the $^{113}$Cd (nuclear spin $I=1/2$, natural abundance 12.26 \%, and gyromagnetic ratio $\gamma/2\pi=9.028$ MHz/Tesla) NMR spectra and spin-lattice relaxation rate (1/$T_{1}$) were performed as a function of temperature in a fixed field magnet ($H$ = 93.9543 kOe) with a room-temperature bore, using standard pulse NMR techniques. The temperature was varied in the range 4-300 K using an Oxford continuous flow cryostat.

\section{Results and Discussions}
\subsection{Crystal structure from XRD}
Rietveld refinement of the collected XRD pattern [Fig. 1(a)-(I) for $T$ = 300 K] ensures nearly single phase with monoclinic $C$2/$c$ space group for Ba$_3$CdIr$_2$O$_9$, except only very small proportion ($<$ 1\%) of unidentified nonmagnetic impurity phase [marked by asterisks in Fig. 1(a)-(I)]. Further, the Energy dispersive X-Ray analysis ensures that the sample is chemically homogeneous and stoichiometry is nearly retained in the target composition, {\it i.e}., Ba:Cd:Ir has been found to be very close to 3:1:2, within the given accuracy of the measurement. Similar to the $M$ = Ca and Sr cases~\cite{Nagprb} the refined crystal structure [Fig. 1(b)] of the Cd compound comprises a larger unit cell, compared to the higher symmetry $P$6$_3$/$mmc$ space group of $M$ = Zn and Mg counterparts~\cite{Nagprl,Nagprb}, in order to accommodate the bigger sized Cd cation. Further careful analysis strongly refutes any Cd/Ir anti-site disorder due to significant ionic size mismatch between Cd$^{2+}$ and Ir$^{5+}$. Like 6$H$ hexagonal perovskites, the Ir ions of this structure form Ir$_2$O$_9$ dimers [Fig. 1(e)], while Cd ions occupy the isolated octahedral sites sharing corners with the Ir$_2$O$_9$ dimers on either side of the $c$-axis. The refined lattice parameters , atomic coordinates, site occupancies along with the goodness factors are summarized in Table-1. Next, we have performed low temperature XRD measurements and consequently, the refined XRD data at $T$ = 4 K is displayed in Fig. 1(a)-(II). Excepting the usual lattice contraction with decreasing temperature, the structure and symmetry remained the same. However, one important observation has been a clear reduction in coordination asymmetry around Ir at 4~K, {\it i.e.} the difference between two Ir-O2 bond lengths has got substantially reduced (see Table-2).
\begin{figure*}
\end{figure*}
Due to lower symmetry monoclinic space group, Ba$_3$CdIr$_2$O$_9$, like Ca/Sr analogues, has 5 different O sites [Fig. 1(e) and Table-1]. This causes six different Ir-O bond lengths, and thus, multiple dissimilar Ir-O-Ir bond angles which should modify the Ir-O-Ir magnetic exchange pathways in this compound. The IrO$_6$ octahedral units within the Ir$_2$O$_9$ dimers are also rotated with respect to each other as a result of monoclinic distortion, as is evident from Fig. 1(c), where O ions are not aligned on top of each other along the $c$-axis. The effect of local non-cubic crystal field ($\Delta^{\text{NC}}_{\text{CFE}}$) around each Ir ion due to five types of oxygen and also monoclinic symmetry driven stronger rotational distortions within IrO$_6$ octahedra units should lift the $t_{2g}$ degeneracy completely for this compound, as for the Ca and Sr ones~\cite{Nagprb}. On top of it, two interesting aspects in the context of BCIO structure are as follows:

i) The intradimer Ir-Ir distance (2.75 \AA) of the Cd compound closely matches with the Mg analogue but is longer than the Ca/Sr counterparts~\cite{Nagprl}. This might cause weakening in intradimer Ir-Ir exchange interactions, compared to the Ca/Sr systems.

ii) The Ir$_2$O$_9$ dimers create almost equilateral triangular units in the $a$-$b$ plane and along $c$ axis [see Fig. 1(c)-(d)], giving rise to several triangular antiferromagnetic interactions. The more symmetric triangular units generate higher degree of frustration in the title compound, exactly like Zn and Mg counterparts~\cite{Nagprl} but unlike the Ca and Sr analogs~\cite{Nagprb} where higher asymmetry in the Ir-triangles reduces frustration.

Therefore, the magnetic ground state in Ba$_3$CdIr$_2$O$_9$ resembles more with the $M$ = Zn and Mg compounds, even though the $M$ = Cd compound possesses an overall lattice symmetry like the Ca, Sr counterparts.



\subsection{Ir-oxidation state and Electronic property}
Estimation of the iridium valence is of central importance in understanding the magnetic ground state of these 5$d$ iridates, as the Ir$^{4+}$/Ir$^{6+}$ species are magnetic~\cite{Kim1,sriro3prb,EJIC_Ir6+,lzioprb} while the Ir$^{5+}$ ion should be ideally non magnetic ($J$ = 0). Therefore, to confirm the oxidation state of Ir in our Ba$_3$CdIr$_2$O$_9$ compound, Ir 4$f$ core level XPS spectrum was collected and fitted using a single spin orbit split doublet [Top left inset to Fig. 2(a)]. The energy positions of 4$f_{7/2}$ (63.10 eV) and 4$f_{5/2}$ (66.15 eV) doublets along with their spin orbit separation around 3.05 eV, confirm pure 5$^+$ charge state of Ir in this compound~\cite{Nagprl,Nagprb,Kenedy,smxpsmrx}. In such a scenario, without the effect of SOC this system should become metallic in a conventional band picture~\cite{Kim1}. To verify this, valence band photoemission spectrum was further collected as shown in the Fig 2(a). As displayed in the top right inset of Fig. 2(a), absence of the density of states right at the fermi level, confirms the insulating behavior of this material. To further check the gapped nature of this compound electrical resistivity has been measured using a standard four probe technique. The temperature dependence of electrical resistivity also supports the insulating behaviour in the entire measured temperature range with $128.5$ Ohm-cm at room temperature, and by 120 K it exceeds 10$^5$ Ohm-cm of resistivity. Although insulating, the resistivity does not follow a thermally activated exponential behaviour $\rho=\rho_0\exp(E_p/k_BT)$ as expected for a simple semiconductor~\cite{Yoshii}. Rather it can be fitted by Mott's variable range hopping (VRH) model in two dimension with $\rho = \rho_0\exp(T_0/T)^{1/3}$.

\subsection{Magnetic Susceptibility}
We have further examined the bulk $dc$ magnetic behaviour of this material. Temperature dependence of both zero field cooled (ZFC) and field cooled (FC) dc magnetic susceptibilities ($\chi$ vs T) have been recorded in the temperature range of 2-300 K, as illustrated in Fig. 3(a). The nearly featureless paramagnetic-like susceptibility curve in the entire temperature range suggests no sign of magnetic long-range ordering in this compound, consistent with several other $d^4$ iridates~\cite{Nagprb,Nagprb2}. In addition, a kink-like feature  is seen at around 50 K. An anomaly at about 50 K is sometimes present due to oxygen condensed on the sample. In the present case, we flushed the sample space with helium gas at 350 K several times but the anomaly still remained. Therefore, there is a possibility that the feature is intrinsic and appearing due to short-range magnetic interactions present in the system. Further, the ZFC-FC divergence starts to develop from $\sim$ 250 K in lower applied fields (data not shown in the figure) and gets completely suppressed under the application of higher magnetic fields, probably indicating the presence of a small fraction of frozen spins~\cite{Okamoto,Dey2} in this compound. Magnetic field dependence of $dc$ magnetization ($M$ versus $H$) has also been measured at 2 K and 300 K and are shown in Fig. 3(b). The absence of any coercivity and remnant magnetization in the 2 K $M$ vs $H$ data strongly refutes hysteresis like behavior, and consequently, possibility of ferromagnetic interactions in this sample. The observed slight nonlinearity at very low fields has been commonly found in several $d^4$ iridate system~\cite{Marco,Dey,ownslio29}. In all these works, the similar nonlinear $M$ vs $H$ curves (not hysteresis) did not represent any ferromagnetic components either.



We have analyzed the susceptibility (in an applied field of 10 kOe) data using the Curie-Weiss (C-W) equation $\chi=\frac{C}{T-\Theta_{\text{CW}}}+\chi_{0}$ ($C$ is the Curie constant while $\Theta_{\text{CW}}$ and $\chi_{0}$ represent the Curie-Weiss temperature and the temperature independent susceptibility, respectively). Note that given the weak $T$-dependence of the susceptibility, there is always some uncertainty about the obtained Curie-Weiss fitting parameters which depend on the temperature range of fitting and also on the applied magnetic fields, similar to other Iridate quantum spin liquids ~\cite{Nagcw}. The key point here is to have an applied magnetic field which should be strong enough to suppress unsaturated paramagnetic components while retaining all existing short range correlations. Under this consideration the 10 kOe $dc$ susceptibility versus temperature data [1/$(\chi-\chi_{0})$ versus $T$ plot, shown in the inset of Fig. 3(a)] appears to be most reasonable for fitting. Our fitting yields a $\Theta_{\text{CW}}$ value of around -21 K and an effective magnetic moment $\mu_{eff}$ $\sim$ 0.3 $\mu_B$/Ir, a closer analogy to the 6$H$ Ba$_3$ZnIr$_2$O$_9$ compound~\cite{Nagprl}. This small finite value of magnetic moment on individual Ir$^{5+}$ site could develop due to intersite real Ir-Ir hopping via Ir-O-Ir paths within Ir$_2$O$_9$ dimers causing delocalization of holes and thus, deviating from a perfect individual atomic $J$ = 0 arrangement~\cite{Nagprb}. Also, local non cubic crystal distortions around the IrO$_6$ octahedral unit may cause mixing of $J$ = 0 singlet with $J$ = 1 triplet state~\cite{Chen,Liu}, and consequently give rise to excitonic moments. The moderate negative value of $\Theta_{\text{CW}}$ suggests antiferromagnetic interaction between the Ir$^{5+}$ moments. Such AFM correlations among the Ir$^{5+}$ moments within the triangular network [Fig. 1(c)] of Ba$_3$CdIr$_2$O$_9$ compound gives rise to magnetic frustration which possibly prevents the system to order. Absence of magnetic ordering with a negative $\Theta_{\text{CW}}$  has been considered as one of the signatures of QSL~\cite{Okamoto,Dey2}. In this context our Ba$_3$CdIr$_2$O$_9$ compound might be proposed as a potential QSL candidate, {\bf like $M$ = Zn and Mg counterparts~\cite{Nagprl,Nagprb}.} Clearly, $dc$ magnetic susceptibility of Ba$_3$CdIr$_2$O$_9$ is quite different from $M$ = Sr and Ca analogues, which exhibit weak ferromagnetic upturn and a dimer like feature respectively along with strong magnetic interactions~\cite{Nagprb}.
\subsection{Heat capacity}
The magnetic frustration of a disordered material is expressed by the amount of magnetic entropy retained within the system at very low temperatures. To further check the magnetic ground state and also to explore the nature of magnetic excitations in Ba$_3$CdIr$_2$O$_9$, we have performed heat capacity measurements ($C_p$ versus $T$) on this sample at zero field and several applied magnetic fields, and the obtained results are displayed in Fig. 4(a). Absence of any sharp $\lambda$-like anomaly in all the measured $C_p$ versus $T$ data [Fig. 4(a)] strongly negates existence of long-range magnetic order and/or structural phase transitions. Although a weak kink-like feature is present at $\sim$ 50 K in the temperature dependent {\it dc} susceptibility [Fig. 3(a)] for this sample, absence of any corresponding signature in heat capacity suggests very feeble magnetic interactions. But, a weak hump-like feature is seen below $\sim$ 10 K in the $C_p$ versus $T$ data, which gets shifted towards higher $T$-values gradually with increasing magnetic field [Fig. 4(a)]. This suggests a two-level Schottky anomaly, possibly arising from some proportion of paramagnetic centers~\cite{Dey2,ownslio28}. The total $C_p$ is thus modeled as the sum of three contributions, namely, the lattice part ($C_{lattice}$), the magnetic contribution ($C_M$) from correlated magnetic moments, and the two-level Schottky anomaly ($C_{sch}$).

In absence of a suitable nonmagnetic analogue, $C_{lattice}$ for this sample was extracted after fitting the $C_p$ data in the $T$ range 70-300 K with the Debye-Einstein equation assuming one Debye term and two Einstein terms. The fitted curve was then extrapolated down to the lowest measuring $T$ [inset to Fig. 4(a]) and taken as the $C_{lattice}$. After subtracting $C_{lattice}$ from the total $C_p$, finally the heat capacity of the sample is left out with ($C_M$ + $C_{sch}$). Now, in order to get the intrinsic magnetic heat capacity $C_M$, `two-level Schottky anomaly' analysis has been performed taking the following strategy: first we subtract the zero field $C_p$ data, {\it i.e.}, $C_p$($H$ = 0) from those in the applied magnetic fields, {\it i.e.}, $C_p$($H$ $\neq$ 0). Consequently, the temperature dependence of this difference ($\Delta$$C_{p,mag}$/$T$), as illustrated in Fig. 4(b), has been fitted with two-level Schottky anomaly equation,
\begin{equation}
\frac{\Delta C_{p,mag}}{T} = \frac{f}{T}[C_{sch}(\Delta(H \neq 0)) - C_{sch}(\Delta(H = 0))]
\end{equation}
where, $f$ is the percentage of paramagnetic centers in the sample; $C_{sch}$($\Delta$) and $C_{sch}$($\Delta_0$) are the Schottky contributions to the specific heat at applied magnetic fields and zero field respectively, while $\Delta$($H$) is the Zeeman splitting in applied magnetic field and $\Delta_0$ represents energy separation between the two levels at $H$ = 0. The $C_{sch}$($\Delta$) is further defined as,
\begin{equation}
C_{sch}(\Delta) = R(\frac{\Delta}{k_BT})^2\frac{\exp(\Delta/k_BT)}{(1 + \exp(\Delta/k_BT))^2}
\end{equation}
The corresponding Schottky fits (shown in Fig.4(b)) yield $f$ $\approx$ 0.5-0.8\%, indicating very small fraction of isolated magnetic centers in this sample. The two-level Schottky gap $\frac{\Delta}{k_B}$ follows a linear relation with applied magnetic field $\Delta$ = $g$$\mu_B$$H$ [inset of Fig.4(b)], consistent with the scenario of free spin Schottky anomalies~\cite{Dey2,ownslio28}. The estimated $g$ value from this linear fit is $\approx$ 1.42.
\par
Finally, the magnetic specific heat $C_M$ is plotted against temperature after eliminating both the lattice and Schottky parts from the measured total $C_p$, and the results are shown in Fig. 4(c) for four different magnetic fields. $C_M$ displays field-dependence at low temperatures ($<$ 10 K), suggesting short-range magnetic interactions in the system. A broad maximum is observed in the $C_M$($T$) data [Fig. 4(c)] at around 25 K which remains unchanged with the application of magnetic fields, supporting frustrated nature of magnetism in this compound as mentioned in the context of some 4$d$ and 5$d$ transition metal based spin liquids~\cite{ownsnio31,Okamoto,ownsnio55}. Unlike in charge insulators, $C_M$ reveals a finite $T$-linear component at very low temperatures for the higher applied magnetic fields, shown in the inset of Fig. 4(c). This points towards gapless nature of the spin density of states or metal-like spinon Fermi surface in this compound~\cite{ownsnio31,Norman,Shen,Nagprb2}. It should be noted that with decreasing field, the temperature range of linear $C_M$ verses $T$ is reduced [inset to Fig. 4(c)], and therefore, any perceptible linear dependence of $C_M$ is missing for both $H$ = 0 and 10 kOe fields. The magnetic entropy change $S_M$ has been obtained by integrating the zero field $\frac{C_M}{T}$ with $T$, as indicated in Fig. 4(d). The percentage of magnetic entropy released is $\approx$ 40\% out of a maximum of $R$$\ln$(2$J$ + 1) with $J$ = 1 corresponding to the hypothetical triplet state. The magnetic entropy release by Ba$_3$CdIr$_2$O$_9$ is clearly more than the Zn and Mg compounds but lesser than the Ca, Sr systems~\cite{Nagprl,Nagprb}, placing it in between the two ends of the 6$H$ hexagonal perovskite family.
\subsection{$^{113}$Cd NMR}
Nuclear magnetic resonance is a useful local probe of magnetism in materials. General experience in insulating oxide-type materials shows that, in the bulk magnetic susceptibility, even small extrinsic Curie-like contributions can overwhelm the intrinsic terms, especially at low-temperatures. In such a situation, the shift of the NMR resonance and its dependence on temperature (in favorable cases where the hyperfine coupling is significant) tracks the intrinsic spin susceptibility. With this motivation, we have measured $^{113}$Cd NMR spectra from 4 K to 300 K. Some representative spectra are shown in Fig. 5(a). The lineshape is not a single symmetric line but rather has additional peaks on either side of the main line. Whereas susceptibility and/or hyperfine coupling anisotropy can give rise to asymmetric lineshapes, this does not appear to be the cause in the present case. Another possibility is the presence of different local environments which is ruled out by the unique crystallographic site for Cd. Some extent of Cd/Ir anti-site disorder could be present though this is less likely due to the ionic size difference between the two. Then one possibility of lineshape anisotropy is due to the distribution of magnetic environments for the $^{113}$Cd nuclei, caused by rotational and tilting distortions of the IrO$_6$ octahedra~\cite{b3yir2o9,b2ymoo6}. The shift of the central line [with respect to $^{113}$Cd reference line, as indicated in Fig. 5(a)] at high-temperatures is about 405 ppm. With decreasing temperature, we find that the main resonance line remains nearly unshifted with temperature. The additional resonance lines (at higher and lower frequencies) also do not shift with temperature.
One can notice some broadening with decreasing temperature and one can see that below about 20 K the additional peaks merge with the main resonance line. Therefore, below $T$ = 20 K, the gradual smearing out of the lineshape-anisotropy with lowering temperature might then imply that the local environments progressively become more homogeneous as a result of weakened local distortions with decreasing temperature~\cite{b3yir2o9,b2ymoo6}, which is consistent with the findings of temperature dependent XRD experiments [Fig. 1(a)]. Now what is the implication of the temperature-independent shift that we observe? In case there is a (transferred) hyperfine coupling between the Cd nuclei and the Ir local moments, the $^{113}$Cd NMR shift ($^{113}K$) should track the spin susceptibility. Note that the bulk susceptibility has a weak Curie-Weiss temperature dependence at low temperatures. This might imply that the observed NMR $K$-shift is just a chemical shift and that there are no local moments. Another possibility is that there are indeed Ir-local moments present (as suggested by the bulk susceptibility) but the hyperfine coupling is very weak as Cd nuclei senses the Ir-local moments through overlap with Ir 5$d$ orbitals via oxygen. The broadening observed at low temperatures, then, definitely comes from a dipolar interaction with the Ir moments.

Next, we present the variation of the $^{113}$Cd NMR spin-lattice relaxation rate 1/$T_{1}$ with temperature. This was determined from a single exponential fit of the recovery of the $^{113}$Cd longitudinal nuclear magnetisation following a saturating pulse sequence. Interestingly, the data  [Fig. 5(b)] show a linear 1/$T_{1}$ versus $T$ dependence upto 150 K. In conventional metals/fermi liquids a Korringa behaviour is expected with $K^{2}T_{1}T=S=\frac{\hbar}{4\pi k_{B}}(\frac{\gamma_{e}}{\gamma_{n}})^{2}$, where $K$ is the NMR Knight shift; $\gamma_{e}$ and $\gamma_{n}$ are the electronic and nuclear gyromagnetic ratios. In case of a metallic $T$-independent susceptibility, a linear 1/$T_{1}$ with $T$ is
expected whereas in semiconductors/non-magnetic insulators, a stronger decrease of 1/$T_{1}$ with $T$ is expected. For our data, we find that $K^{2}T_{1}T/S$ is close to 1, implying a possible fermi liquid state in our sample. But this is clearly refuted from our resistivity data. On the other hand, spin liquids have quasiparticles with spin degrees of freedom but no charge degrees of freedom. Hence, the linear 1/$T_{1}$ versus $T$ [in the temperature range $T$ = 4-150 K, shown in Fig. 5(b)] is suggestive of a spinon fermi surface. The absence of any significant $T$-dependence of $^{113}K$ in the present case (though the bulk susceptibility is Curie-Weiss-like) has therefore to be attributed to a weak hyperfine coupling together with weak intrinsic Ir-local moments. After all, the effective magnetic moment per iridium as inferred from the Curie constant (see magnetization discussion in Section-C) is only about 17\% of that expected for spin-1/2 paramagnetic Ir$^{4+}$ moments.

\section{Summary and Conclusion}
In conclusion, unlike the iso-structural 6$H$ analogue Ba$_3$$M$Ir$_2$O$_9$ ($M$ = Ca, Sr) compounds, Ba$_3$CdIr$_2$O$_9$ reveals closer realization to the elusive $J$ = 0 state. The small spin-orbital moments per Ir likely develop through real intersite hopping and noncubic crystal distortions. Despite moderate AFM interactions between these Ir$^{5+}$ moments, our combined ${}^{113}$Cd NMR, heat capacity, and dc susceptibility results discern the absence of magnetic ordering down to at least 2 K, invoking the possibility of QSL-like ground state for this compound. Both the XPS valence band spectra and electrical resistivity measurements suggest gapped electronic structure of this compound, which mimics the dominance of SOC, as in other 5$d$ iridates. On the other hand, within the spin sector, the linear temperature dependence of ${}^{113}$Cd NMR spin-lattice relaxation rate and also the linear magnetic heat capacity at low $T$ are suggested to arise from the gapless nature of spin excitations within this material. Further measurements at even lower temperatures would be extremely useful to establish the absence of static magnetic moments in the ground state.

\section{Acknowledgement}
MSK thanks UGC, India for fellowship. AB and VK thank CSIR, India for fellowship. SR thanks Department of Science and Technology (DST) [Project No. WTI/2K15/74] and TRC, IACS for support. Authors thank Indian Institute of Technology, Bombay and Indian Association for the Cultivation of Science, Kolkata for support in research. Authors also thank Prof. Subham Majumdar of IACS for useful discussions.

\newpage

\begin{table}
\caption{Ba$_3$CdIr$_2$O$_9$ refined structural parameters (space group: $C2/c$, $\alpha$ = $\gamma$ = 90$^{\circ}$, $\beta$ = 91.352$^{\circ}$), $a$ = 5.8990\AA, $b$ = 10.19745\AA,, $c$ = 14.74057\AA, $R$$_{p}$ = 20.1, $R$$_{wp}$= 16.3, $R$$_{exp}$=6.72, and $\chi$$^{2}$=5.88}
\resizebox{8.8cm}{!}{
\begin{tabular}{c c c c c c c}
\hline Atom & Site & occupancy & $x$ & $y$ & $z$ & $B$({\AA})$^2$ \\\hline
Ba(1) & 4e & 0.5 & 0 & -0.0042(6) & 1/4 & 0.159(2)\\
Ba(2) & 8f & 1 & 0.0142(3) & 0.3338(4) & 0.0924(3) & 0.396(2)\\
Cd & 4a & 1 & 0 & 0 & 0 & 0.181(4)\\
Ir & 8f & 1 & -0.0167(2) & 0.0043(3) & 0.8439(5) & 0.306(4)\\
O(1) & 4e & 0.5 & 0 & 0.5304(6) & 1/4 & 1.1(3)\\
O(2) & 8f & 1 & 0.2834(6) & 0.2542(8) & 0.2411(6) & 1.1(3)\\
O(3) & 8f & 1 & 0.0364(5) & 0.8273(3) & 0.1012(4) & 1.1(3)\\
O(4) & 8f & 1 & 0.2985(3) & 0.0843(2) & 0.0752(8) & 1.1(3)\\
O(5) & 8f & 1 & 0.7625(6) & 0.0783(4) & 0.0932(5) & 1.1(3)\\
\hline
\end{tabular}
\label{$1$}
}
\end{table}

\newpage

\begin{table*}
\caption{Ba$_3$CdIr$_2$O$_9$ refined structural parameters comparison between room temperature (300K) and at low temperature (4K)}
\resizebox{16cm}{!}{
\begin{tabular}{|c|c|c|c|c|c|}
\hline & & & & & \\
 Temperature(K) & Crystal structure & Bond-lengths & Intra-dimer Ir-Ir (\AA) & Inter-dimer Ir-Ir along $c$ axis(\AA) & Inter-dimer Ir-Ir along $a-b$ plane (\AA) \\
  & & & & & \\\hline
  & S.G: C2/$c$ &  Ir-O1 (\AA) = 1.951 & & & \\
  & a = 5.8990 (\AA) &  Ir-O2 (\AA) = 2.184, 1.998 &   &  & \\
  300 & b = 10.1974 (\AA) &  Ir-O3 (\AA) = 1.840 & 2.75 & 5.73, 5.73, 5.76  & 5.890, 5.899 \\
  & c = 14.7405 (\AA) &  Ir-O4 (\AA) = 1.915 &&& \\
  & $\beta$ = 91.352$^{\circ}$ &  Ir-O5 (\AA) = 1.984 &&& \\
  \hline
  & S.G: C2/$c$ &  Ir-O1 (\AA) = 1.946 &&& \\
  & a = 5.8752 (\AA)&  Ir-O2 (\AA) = 1.987, 1.997 &&& \\
  4 & b = 10.1615 (\AA) &  Ir-O3 (\AA) = 1.835 & 2.75 & 5.73, 5.73, 5.71  & 5.870, 5.875 \\
  & c = 14.7145 (\AA) &  Ir-O4 (\AA) = 1.905 &&& \\
  & $\beta$ = 91.701$^{\circ}$ &  Ir-O5 (\AA) = 1.981 &&& \\

\hline
\end{tabular}
\label{$2$}
}
\end{table*}

\begin{figure*}
\caption{(color online) Rietveld refined XRD pattern of Ba$_3$CdIr$_2$O$_9$ sample at 300K (a-I)and 4K (a-II); $``*"$ denotes unidentified nonmagnetic impurity phase; (b)Refined crystal Structure, indicating exchange interaction paths; (c)Edge-shared frustrated Ir-triangular network in $a$-$b$ plane; (d)Frustrated Ir-triangular network along $c$ axis; (e) Ir$_2$O$_9$ dimer unit consisting of two face-shared  IrO$_6$ octahedra, and 5 different types of oxygen}
\end{figure*}

\begin{figure*}
\caption{(color online) (a) Valance band photoemission spectra for Ba$_3$CdIr$_2$O$_9$; Top right inset: expanded view of the valance band near the Fermi level; Top left inset: Ir 4$f$ core level XPS spectrum (shaded black circles) along with the fitting (red solid line) (b) Temperature dependent zero field electrical resistivity variation, Inset: Respective Mott VRH fitting in 2-$D$}
\end{figure*}

\begin{figure}
\caption{(color online)(a)Temperature dependent $dc$ susceptibility variations during zero-field-cooled (open blue circles) and field-cooled (shaded blue circles) protocols; Inset: Temperature dependence of 1/($\chi$-$\chi_0$)is plotted,(b) Field dependence isothermal magnetization at 4 K and 300 K.}
\end{figure}

\begin{figure*}
\caption{(color online) (a)Temperature dependence of total specific heat $C_p$ in low-$T$ region for zero field and also several applied magnetic fields, Inset: Lattice contribution to the zero field $C_p$ data;  (b) Temperature variations of the $T$-divided [$C_p$($H\neq$ 0)- $C_p$($H$=0)] at several magnetic fields (shaded black circles) along with the two level Schottky anomaly fits(solid red lines), Top inset: Magnetic field dependence of the Schottky energy gap($\Delta/K_B$ verses $H$, shaded black circles and the subsequent linear fitting(solid red line);(c) Temperature dependence of the magnetic specific heat $C_M$ at zero field and the applied magnetic fields, inset: Zoomed-in views of the respective low-$T$ linear $C_M-T$ variations; (d)Magnetic entropy loss as a function of temperature.}
\end{figure*}

\begin{figure*}
\caption{(color online) (a) $^{113}$Cd NMR line shapes at several temperatures (solid coloured lines) along with the $^{113}$Cd reference (vertical dotted brown line), (b) Temperature dependence of the NMR spin-lattice relaxation rate (magenta half-filled circles) along with the linear fitting (solid orange line), error bars are also shown with the experimental data.}
\end{figure*}

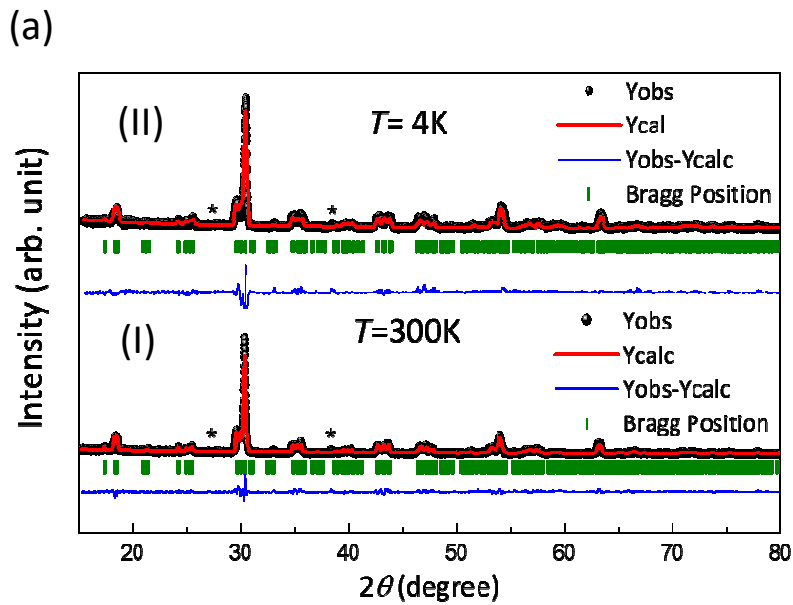
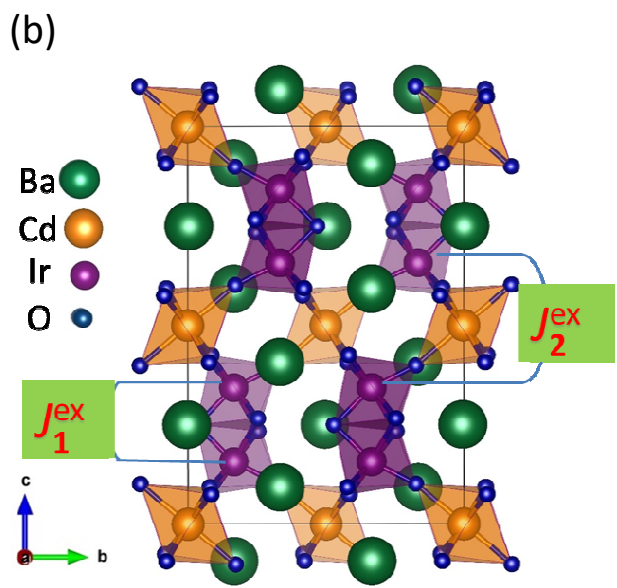
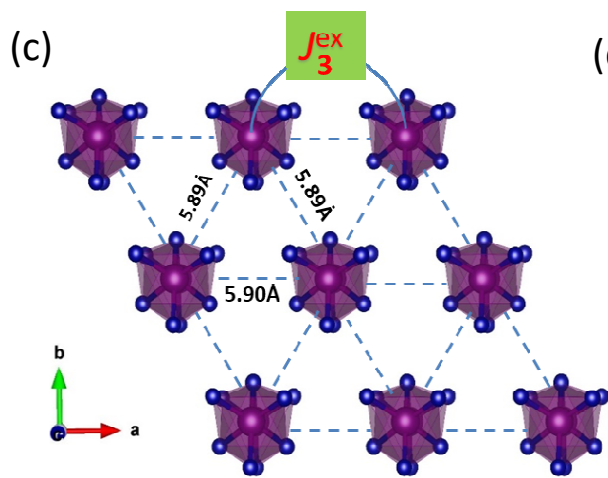
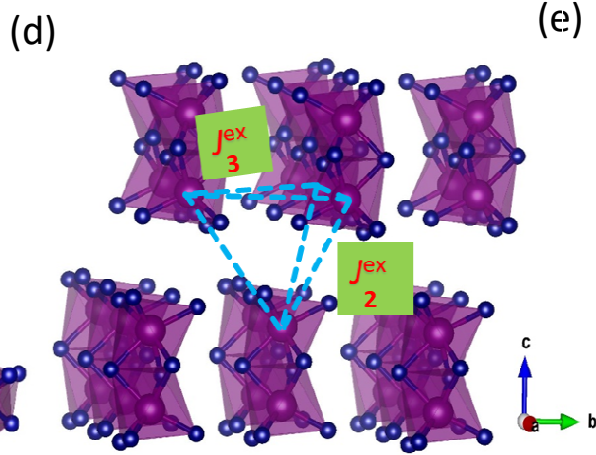
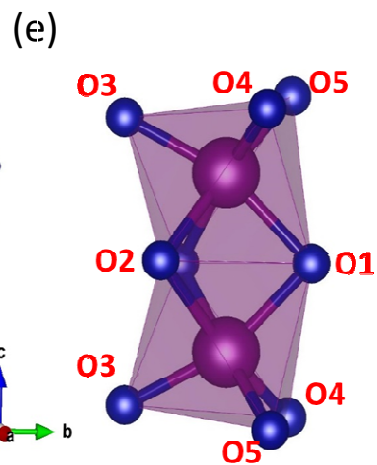
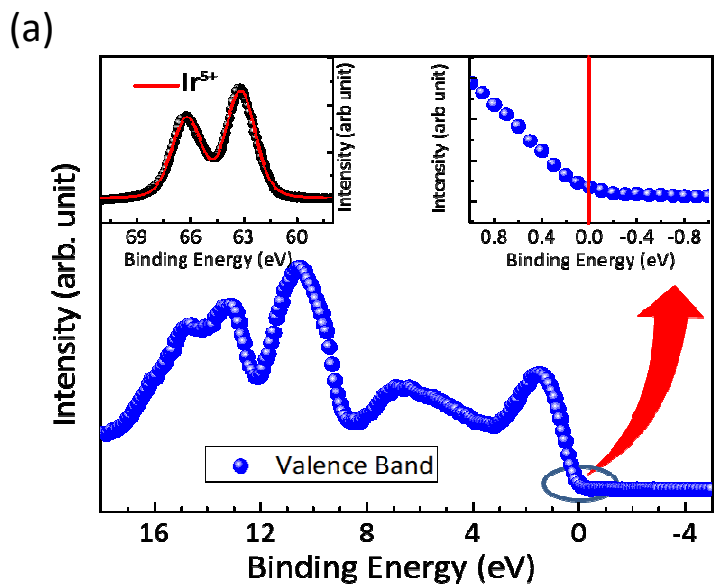
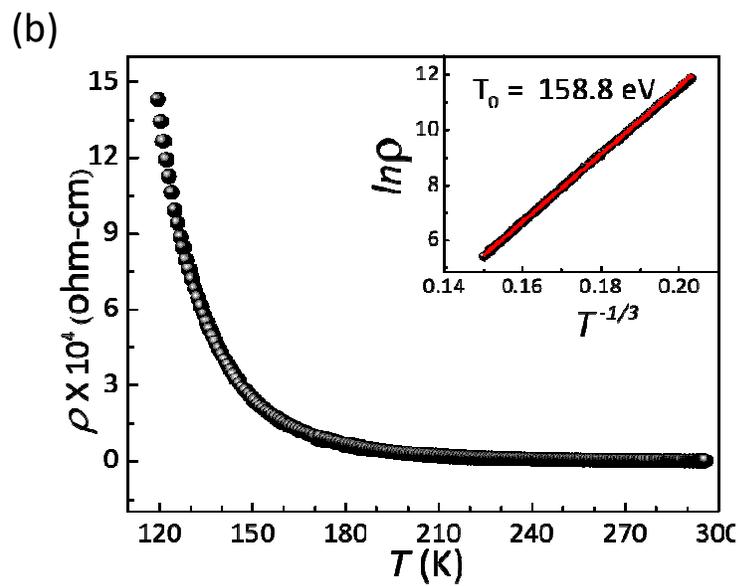
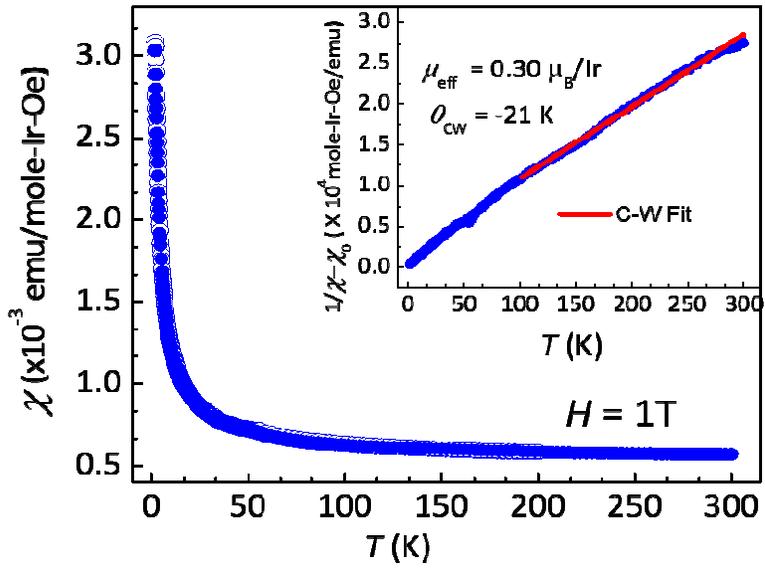
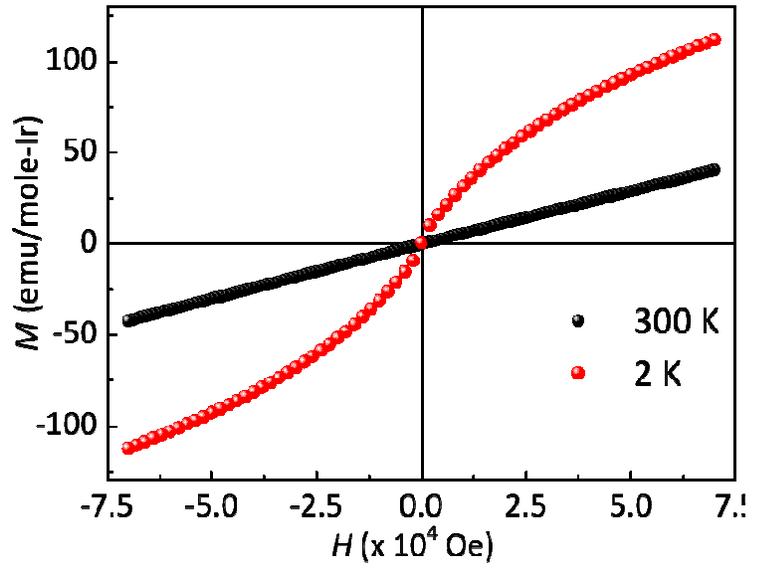
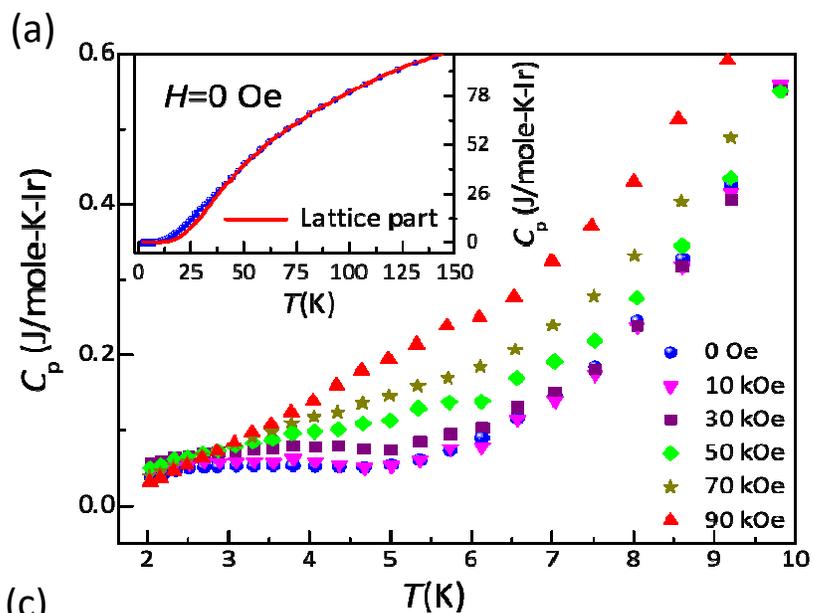
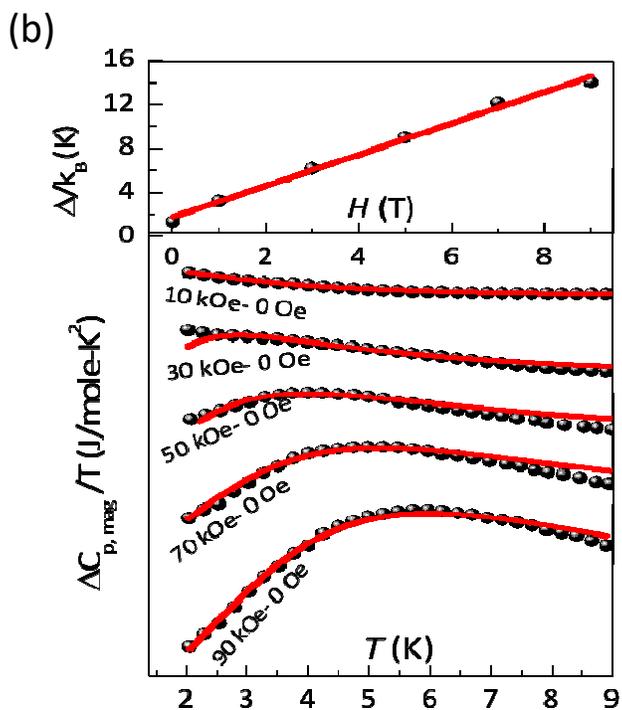
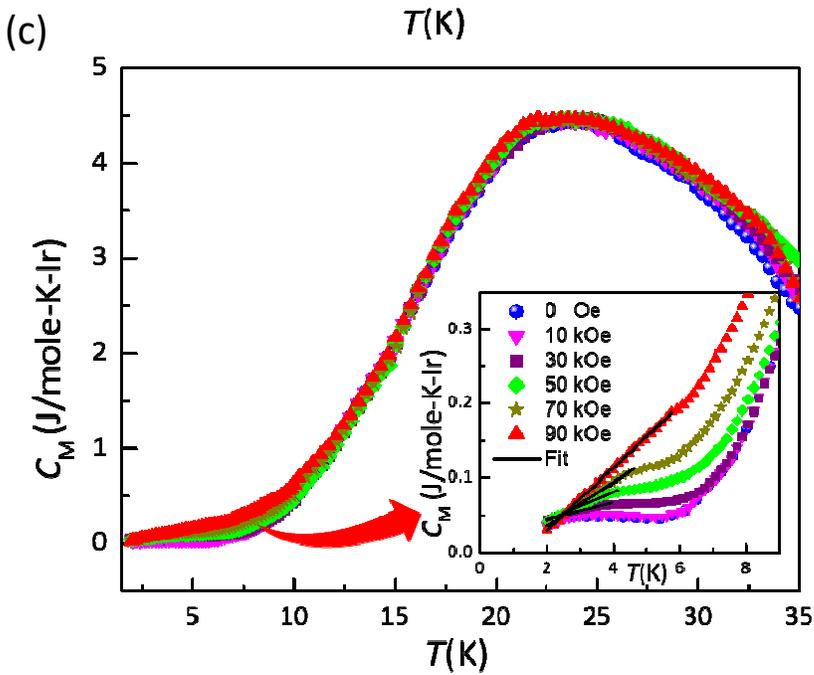
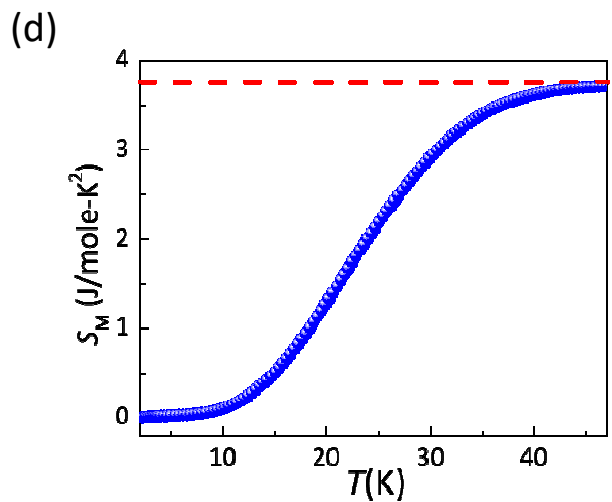
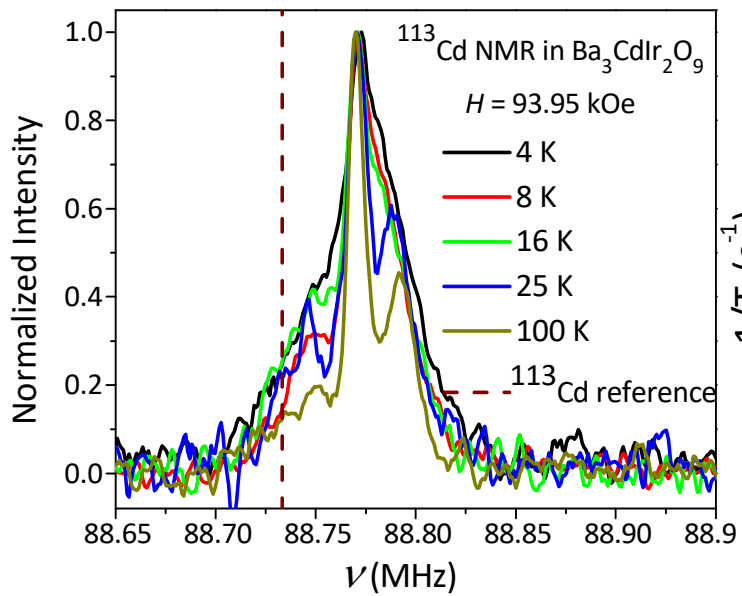
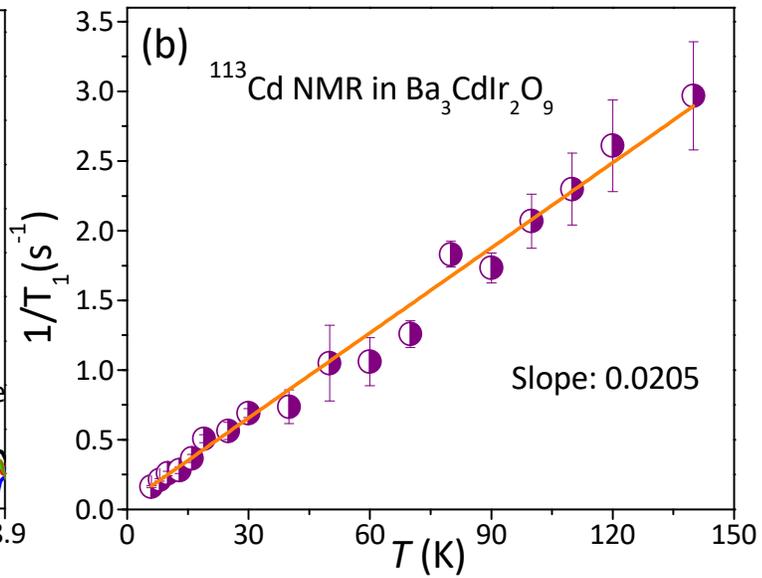
\end{document}